# New low temperature phase in dense hydrogen: The phase diagram to 420 GPa

Ranga P. Dias, Ori Noked, and Isaac F. Silvera

Lyman Laboratory of Physics, Harvard University, Cambridge MA, 02138

In the quest to make metallic hydrogen at low temperatures a rich number of new phases have been found and the highest pressure ones have somewhat horizontal phase lines with increasing pressure around room temperature. We have studied hydrogen to static pressures of $420 \pm 13$ GPa in a diamond anvil cell and down to liquid helium temperatures, using infrared spectroscopy. We observe a new phase at a pressure of $355 \pm 13$ GPa and T=5 K, with an almost vertical slope for the phase line. Although we observe strong darkening of the sample in the visible, we have no evidence that this phase is metallic.

In 1935, over 80 years ago, Wigner and Huntington (**WH**) predicted that solid molecular hydrogen would dissociate and transform to atomic metallic hydrogen (**MH**) if pressurized to 25 GPa (1 Mbar=100 GPa), in the low temperature limit [1]. WH stimulated decades long challenging theoretical and experimental studies of dense hydrogen, with experimental static pressures over an order of magnitude higher than their transition pressure prediction. These studies have not revealed the metallic phase (discussed ahead). Theory for this simple atomic system is challenging due to the low atomic mass with large zero-point energy and motion of the molecules or atoms that is not easily handled. Sophisticated analyses have begun to converge in their predictions. The great challenge for experiment is the modern predicted pressures of 400-500 GPa, required for metallization [2-4]; only recently have such pressures been approached in the laboratory. In this Letter we report on the highest reported static pressures ever achieved on hydrogen, 420 GPa. A new phase is observed at liquid helium temperatures, but is not metallic.

MH is predicted to have spectacular properties such as room temperature (**RT**) superconductivity [5], possible metastability, so that if produced it might exist at ambient conditions [6], and a prediction that the megabar pressure atomic metallic phase may be a liquid at T=0 K [6]. A determination of the phase diagram and equation of state of hydrogen is crucial to our understanding of the structure of the astrophysical gas giants, Saturn and Jupiter [7]. If MH is superconducting and metastable, it would have an important impact on energy transmission, and it would revolutionize rocketry as a remarkably light and powerful propellant [8]. Earlier static experiments to well over 300 GPa have failed to reveal MH [9-13]. Below, we briefly review advances in the study of hydrogen; we then present our studies to 420 GPa and the observation of a new phase. New phases in hydrogen have recently been reported elsewhere [14,15] and will be compared to our observations.

There have been important developments in the experimental and theoretical phase diagram of hydrogen, shown in Fig. 1, where we see a large predicted region (grey) for liquid atomic metallic hydrogen. Although we are confident that MH will exist at ultra high pressures, it is not clear that it will be solid or liquid in the low temperature limit (as indicated in the figure). In this phase diagram a maximum in the melting line was predicted with an extrapolation to low temperature at high pressure [16,17]. The solid part of the red line is a fit to existing experimental data, in fairly good agreement with the theoretical curve; the maximum was first observed by Deemyad and Silvera [18], and confirmed by a number of other measurements [19-22]. A somewhat similar phase diagram exists for deuterium with shifts in some of the phase lines, but in this Letter we focus on hydrogen. In particular we explore the low temperature Pathway I in search of the WH transition and the experimentally observed solid phases along this Pathway.

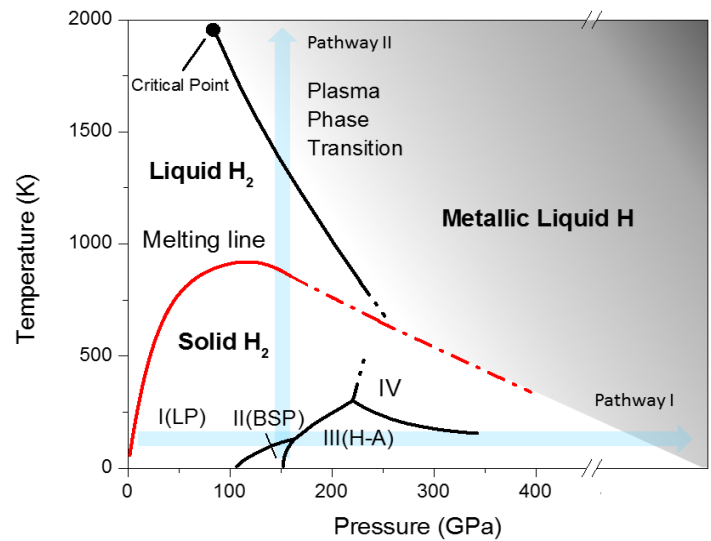

Fig. 1. A current theoretical/experimental phase diagram of hydrogen, showing the melt line with a maximum and a linear extrapolation to low temperature, as well as solid phases that have been observed, based on the lines for pure para-hydrogen (LP, BSP, and H-A), and the phase line for the plasma phase transition to liquid MH. Two possible experimental pathways to MH, Pathway I and Pathway II, are indicated by arrows.

The observed solid phases at low to intermediate pressures are shown in Fig. 1; four phases have been identified along this pathway. The first three are named LP (low pressure), BSP (broken symmetry phase), and H-A [23-25]. These are the phase names for pure para-hydrogen based on the symmetric and antisymmetric nuclear spin states, which are expected to be preserved until the dissociative WH transition [26]. If the sample has a mixed ortho-para concentration, then there is no translational symmetry in the solids, but structural phase transitions are still observed to occur. In such cases the transitions shift to lower pressures [27], in particular for the BSP phase. Mixed ortho-para samples are named phases I, II, III, respectively [28] to be distinguished from the pure isomer phases. Phases IV and V (V is not shown in Fig. 1) were first observed (but not named) by Eremets and Troyan [11] at pressures of 220 and 270 GPa, respectively, by noting changes in the Raman scattering spectra in hydrogen. Later, Howie et al [12] studied phase IV, also by Raman scattering around RT (and named it IV); they also observed a higher pressure phase at 275 GPa which they called IV' [29]; this appears to be the second phase observed by Eremets and Troyan [11], or phase V, to be discussed later. Phase IV [11,12]) is always a mixed ortho-para phase as it only occurs at elevated temperatures. These phases represent structural changes characterized by orientational-order of the molecules in which the solid remains insulating. Recently, Dalladay-Simpson, Howie, and Gregoryanz (**DHG**) [14] reported a new phase at 325 GPa which they named V. We shall clarify the naming of phases below. In this Letter we find a new high-pressure phase that occurs at low-temperature. At high densities ortho-para conversion should be very rapid [26] so that this new phase is pure para-hydrogen.

In Fig. 1, the thick arrows indicate two Pathways for achieving MH, Pathway I and II. Metallic hydrogen has been achieved along Pathway II at static pressures [30], which is easier to achieve as ultra-high pressures are not required. We show the phase line for the Plasma Phase Transition (PPT), a first-order phase transition from insulating liquid-molecular to liquid-atomic MH. Earlier dynamic shock wave measurements reported a continuous change to metallic hydrogen by measuring conductivity or optical properties at high pressures and typically much higher temperatures than those shown in Fig. 1 [31-38]. The authors of a recent ramp-shock measurement on deuterium [38] suggest that a first-order transition may have been observed, but temperature was not measured, which is important for determining a P-T phase line. The objective of the current experiment was to traverse Pathway I to the highest pressures at temperatures as low as 5 K.

There are two challenges for studying hydrogen at high pressure, besides the strength of the diamond anvils. First, hydrogen is very compressible and the molar volume changes by a factor of ~10 when pressurized to the megabar region so that stable, precision alignment of the DAC is required or the sample blows out of the gasket as pressure is increased. The second challenge arises from hydrogen diffusion. Hydrogen is very reactive and can diffuse into the metallic gasket or diamonds; diamond anvils can embrittle and fail. Diffusion is an activated process and is suppressed at low temperature, but can lead to diamond failure at high P, T, even RT, so in this experiment we do not explore the room temperature region.

Hydrogen was cryogenically loaded in a DAC and cryostat similar to a design described elsewhere [39]. A rhenium gasket confines the sample. Type IIac conic diamonds with 30-micron diameter culet flats were used. At the highest pressures the mean dimension of the sample was 9-10 microns. The diamonds were coated with alumina that acts as a diffusion barrier against hydrogen. We studied hydrogen by infrared spectroscopy, using a Nicolet Fourier Transform infrared interferometer with a thermal source. Our optical system also enabled Raman scattering on the sample. Pressure was measured at lower pressures using the Ruby scale calibration of Chijioke et al [40] and from 150 to 350 GPa using the vibron pressure scale of Zha et al [13] (based on the 2010 diamond Raman scale of Akahama and Kawamura [41]). For higher pressures we measured the diamond Raman line [41]. We round our specified pressures to the nearest 0 or 5 GPa.

Figure 2 shows infrared spectra in the region of the hydrogen vibron lines at various pressures and temperatures. In general, a splitting or discrete shift of excitations as a function of pressure or temperature arises from a change of lattice symmetry and can be used to identify a phase transition. The fundamental vibron spectrum at 150 GPa (~4403 $cm^{-1}$) shows coexisting phases in the transition from phase II to III. Above 355 GPa, at 5 K, we observe an abrupt spectral change (see Fig. 2a). The fundamental vibron disappears, and two new broad absorption lines appear at ~2964 $cm^{-1}$ and ~3299 $cm^{-1}$, as well as a very weak absorption line around ~4322 $cm^{-1}$ (raw data and normalization procedures are presented in the Supplementary Materials, **SM**). Above 385 GPa the high frequency absorption line disappears, while the two broad new peaks remain. There is a similar spectrum at higher temperature, T=82 K, shown in Fig. 2b. These spectra identify a transition to a new phase and the boundaries of a phase line, shown in Fig. 3.

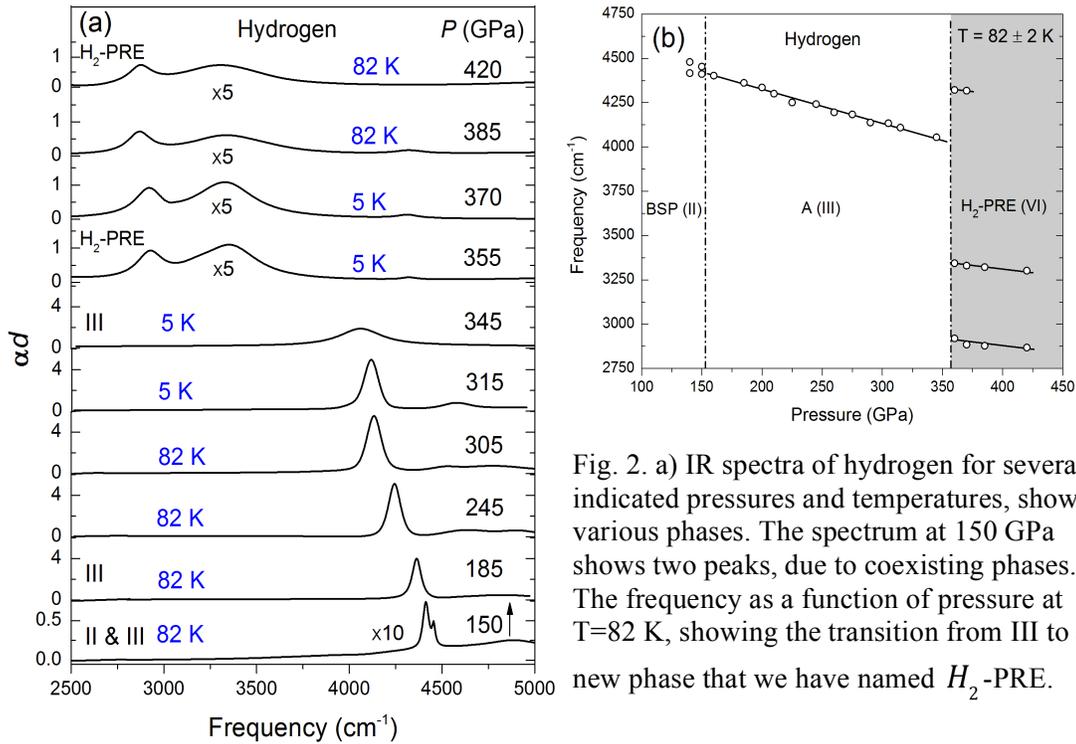

Fig. 2. a) IR spectra of hydrogen for several indicated pressures and temperatures, showing various phases. The spectrum at 150 GPa shows two peaks, due to coexisting phases. b) The frequency as a function of pressure at T=82 K, showing the transition from III to a new phase that we have named $H_2$-PRE.

The accuracy of theoretical predictions of the phase diagram of hydrogen has rapidly improved in the past several years. It was recognized that a straightforward application of density functional theory (DFT) gave unreliable results and various forms of DFT were compared [42]. Recently theorists have focused on the use of diffusion Monte Carlo (DMC); this overcomes the deficiencies of DFT, and with corrections for zero-point energy and anharmonic effects the predictive powers have made great gains. For example, the phase line for the PPT agrees with experiment to within 25 GPa for hydrogen [43] and P-T values for phase IV are in reasonable agreement with experiment [44]. Two recent DMC calculations of the high pressure transition to MH [3,4] predict the metallization pressure to be around 400 GPa with the sample transforming from phase III to a molecular structure Cmca-12 to an atomic metallic structure $I4_1/amd$ (the Cs-IV structure) with increasing pressure at low temperature [45].

The experimentally observed phase line is consistent with the DFT calculation of a transition to a Cmca-12 phase which was shown to be a quantum phase transition [46], i.e. the phase line rises almost vertically from the T=0 K limit. Unpublished DFT calculations, including quasi-harmonic corrections, predict 3 IR active modes in our frequency range, but different functionals give different mode frequencies, so calculations need refinement. (Ref. to Clay). Thus, the new phase that we have observed may be Cmca-12 [45]. In concurrence with theoretical predictions, we believe that this new phase may precede the transition to solid metallic hydrogen and we name it $H_2$-PRE.

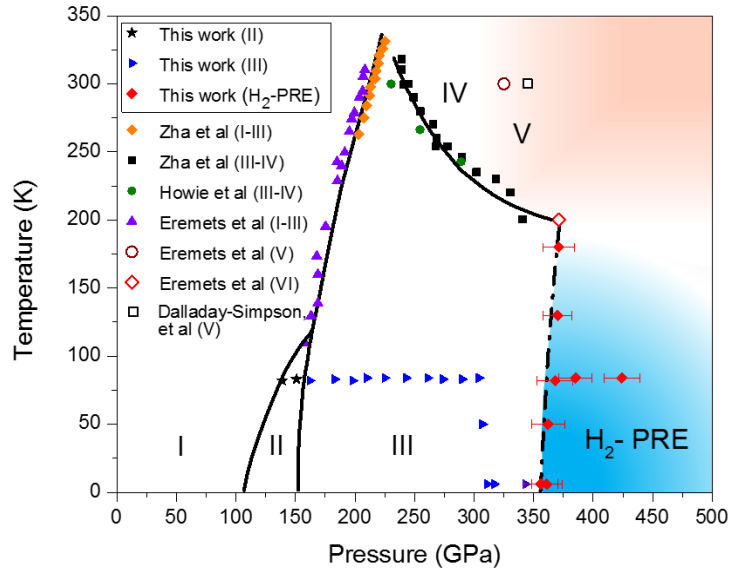

Fig. 3. Phase diagram for hydrogen along Pathway I showing new phase lines. Our data points indicate the thermodynamic pathway followed in the experiment and the new phase observed at low temperatures and 355 GPa. The new phase reported by Eremets et al seems to be on the phase line for the phase we call $H_2$-PRE, so the line (dash-dot) is extended to include this point. A higher temperature point from Eremets et al (open circle) indicates the cross over from mixed (IV/V) to single phase (V); the new phase claimed by Dalladay-Simpson et al is also shown (open square, rescaled to the 2010 pressure scale of Akahama and Kawamura, the same scale used for other displayed measurements). The uncertainties in the points for $H_2$-PRE are statistical from determining the pressure of the diamond phonon; systematic errors from different researchers can be of order $\pm 13$ GPa. The absence of phase IV' is discussed in the text and presented in Table I.

We show the extended phase diagram (from Fig. 1) of hydrogen along Pathway I in Fig. 3. While our letter was under preparation, a study of hydrogen by Eremets, Troyan, and Drozdov (**ETD**), using Raman scattering, appeared on the arXiv [15]. They observed the disappearance of a low frequency Raman mode at 360 GPa and T~200 K, but did not study other spectral regions for the simultaneous appearance of new vibron modes. ETD state that their study is at a preliminary stage and are working on further verification of the new phase, which they name VI. If we extend our phase line in Fig. 3, it appears that their point falls on this line. Due to expected rapid ortho-para conversion, the para concentration varies along this line, depending on temperature. Thus, in the tradition of naming of phases, the pure phase is $H_2$-PRE, while the higher temperature mixed phase is VI, assuming that it is confirmed in further studies by ETD. Both we and ETD use the same pressure scale (see Fig. 10 of ETD).

The lower pressure phases, II and III, are quantum phase transitions, appearing in the low temperature limit as pressure is increased. On the other hand, Phase IV only appears around RT and has a relatively weak pressure dependence, as seen in Fig. 3. Magdau and Auckland [47], and others [46,48] simulated such phases and argued that these transitions are entropy driven, so that they only appear at elevated temperatures. Several studies delineating and searching for new phases focused on temperatures around RT [11-14,22,29,49-53]. The study of HD by Dias, Noked, and Silvera [54] returned the studies to low temperatures. They showed that above a threshold pressure, HD dissociates into a mixture of $H_2$, $D_2$ and $HD$. They did not observe a phase similar to IV, but two new high pressure phases named HD-IV* and HD-PRE, with steep phase lines rising from liquid helium temperatures. In this letter on hydrogen we also observe a new phase, with a phase-line rising from low temperature, as was the case for phase lines in HD. We conclude that the ultra high-pressure hydrogens exhibit quantum phase transitions.

In a recent paper DHG [14] claim to observe a new phase at a pressure of 325 GPa, which they named phase V. This is based, to a large extent, on changes of intensity of Raman modes, which is not a signature of a phase transition. They show (their Fig. 3) that phase IV and IV' have a large pressure range (270-320 GPa) where they coexist and solid hydrogen becomes single-phase for higher pressures, phase V. They did not determine the phase lines between IV, IV' and V. ETD give several plausible reasons that could explain the observations of DHG as being a misinterpretation of data for phase IV'. They do not observe the new phase proposed by DHG, and neither do we in the low temperature IR spectra; however, we did not study hydrogen at room temperature in the IR. (For the new phase DHG used the 2004 pressure scale of Akahama and Kawamura [55], rather than their improved 2010 calibration [41]). We provide more evidence of this implied misinterpretation of data by DHG elsewhere [56]. The conclusion is that phase IV and IV' coexist and as pressure is increased phase IV diminishes in favor of IV'. Thus it appears that IV' and V are one and the same phase that begins at ~270 GPa.

There are a number of phases, names of phases, and claims of properties for solid hydrogen that can be confusing; here we hope to clarify the situation. As a guide to Fig. 3, we also track these phases and their names in Table I. Phases I, II, and III, discussed above, are well known and established, so they are omitted from further discussion. We believe that IV' is identical to phase V, so that <u>the name IV' should be dropped.</u> Zha et al [13,50] and Loubeyre et al [52] also studied phase IV by infrared spectroscopy and do not mention V or IV'.

At the highest pressures our sample was dark in reflected and transmitted light, but we could still see faint transmitted light in the visible (see SM). We were unable to make quantitative measurements of visible transmission, but could measure a spectrum in the IR (see SM). Others have observed darkening of high pressure hydrogen [10],[57].

| Reference | Pressure Range (GPa) | | | | Comment |
|---|---|---|---|---|---|
| | 220 | 270-320 see note | 325 | 355-420 | |
| Eremets et al (2011) | (IV) | (V) | | | Parentheses mean unnamed |
| Howie et al (2012) | IV | IV' | | | IV' $\equiv$ V |
| Zha et al & Loubeyre et al (2013) | IV | | | | |
| Dalladay-Simpson et al (2016) | IV | IV' | V | | IV' $\equiv$ V |
| Eremets et al (2016) | IV | V | V | VI | VI observed at ~200 K. |
| This Work | | | | $H_2$–PRE | Observed at 5 K and higher temp. |
| Proposed Naming of Phases | IV | V | V | $H_2$–PRE ,VI | VI $\equiv$ $H_2$–PRE at lower temp. |

Table I. Phases of hydrogen that have been discussed and named in the literature. Note: Phases IV and V are proposed by Eremets et al [15] to coexist in the region 270-320 GPa, leading to the conclusion that phase IV'(named by Howie et al [29]) is the same as phase V. Phase $H_2$–PRE is the name for the new para-hydrogen phase observed at liquid helium temperatures.

99

In conclusion, we have studied solid hydrogen to the highest pressures yet reported in the literature and to liquid helium temperatures. We observed a new quantum phase transition at a pressure of 355 GPa in the low temperature limit, consistent with theory; the phase is not metallic. Our proposed clarification of the naming of the solid phases and pressure regions is summarized in Table I. Pressure uncertainties may be of order ±13 GPa due to use of the diamond scale [58], but we do not think this affects the conclusions reached here.

We thank Mohamed Zaghoo and Rachel Husband for useful discussions. The NSF, grant DMR-1308641, the DOE Stockpile Stewardship Academic Alliance Program, grant DE-NA0001990, supported this research. Preparation of diamond surfaces was performed in part at the Center for Nanoscale Systems (CNS), a member of the National Nanotechnology Infrastructure Network (NNIN), which is supported by the National Science Foundation under NSF award no. ECS-0335765. CNS is part of Harvard University.

**References**

[1] E. Wigner and H. B. Huntington, J. Chem. Phys. **3**, 764 (1935).
[2] J. M. McMahon, M. A. Morales, C. Pierleoni, and D. M. Ceperley, RevModPhys **84**, 1607 (2012).
[3] J. McMinis, R. C. C. III, D. Lee, and M. A. Morales, Phys. Rev. Lett. **114**, 105305 (2015).
[4] S. Azadi, B. Monserrat, W. M. C. Foulkes, and R. J. Needs, Phys. Rev. Lett. **112**, 165501(5) (2014).
[5] N. W. Ashcroft, Phys. Rev. Lett. **21**, 1748 (1968).
[6] E. G. Brovman, Y. Kagan, and A. Kholas, Sov. Phys. JETP **34**, 1300 (1972).
[7] D. J. Stevenson, Journal of Physics: Condensed Matter **10**, 11227 (1998).
[8] J. Cole, I. F. Silvera, and J. P. Foote, in *STAIF-2008*, edited by A. C. P. 978Albequerque, NM, 2008), pp. 977.
[9] C. Narayana, H. Luo, J. Orloff, and A. L. Ruoff, Nature **393**, 46 (1998).
[10] P. Loubeyre, F. Occelli, and R. LeToullec, Nature **416**, 613 (2002).
[11] M. I. Eremets and I. A. Troyan, Nature Materials **10**, 927 (2011).
[12] R. T. Howie, C. L. Guillaume, T.Scheler, A. F. Goncharov, and E. Gregoryanz, Phys. Rev. Lett. **108**, 125501 (2012).
[13] C.-S. Zha, Z. Liu, and R. J. Hemley, Phys. Rev. Lett. **108**, 146402 (2012).
[14] P. Dalladay-Simpson, R. T. Howie, and E. Gregoryanz, Nature **529**, 63 (2016).
[15] M. I. Eremets, I. A. Troyan, and A. P. Drozdov, arXiv:1601.04479 (2016).
[16] S. A. Bonev, E. Schwegler, T. Ogitsu, and G. Galli, Nature **431**, 669 (2004).
[17] I. Tamblyn and S. A. Bonev, Phys. Rev. Lett. **104**, 065702 (2010).
[18] S. Deemyad and I. F. Silvera, Phys. Rev. Lett. **100**, 155701 (2008).

[19]   M. I. Eremets and I. A. Trojan, JETP Letters **89**, 174 (2009).
[20]   N. Subramanian, A. V. Goncharov, V. V. Struzhkin, M. Somayazulu, and R. J. Hemley, Proc. of the National Academy of Sciences **108**, 6014 (2011).
[21]   V. Dzyabura, M. Zaghoo, and I. F. Silvera, Proc Natl Acad Sci U S A **110**, 8040 (2013).
[22]   R. Howie, P. Dalladay-Simpson, and E. Gregoryanz, Nature Materials **14**, 495 (2015).
[23]   I. F. Silvera and R. J. Wijngaarden, Phys. Rev. Lett. **47**, 39 (1981).
[24]   R. J. Hemley and H. K. Mao, Phys. Rev. Lett. **61**, 857 (1988).
[25]   H. E. Lorenzana, I. F. Silvera, and K. A. Goettel, Phys. Rev. Lett. **63**, 2080 (1989).
[26]   I. F. Silvera, J. Low Temp. Phys. **112**, 237 (1998).
[27]   I. F. Silvera, Rev. Mod. Physics **52**, 393 (1980).
[28]   L. Cui, N. H. Chen, S. J. Jeon, and I. F. Silvera, Phys. Rev. Lett. **72**, 3048 (1994).
[29]   R. T. Howie, T. Scheler, C. L. Guillaume, and E. Gregoryanz, Phys Rev B **86**, 214104 (2012).
[30]   M. Zaghoo, A. Salamat, and I. F. Silvera, Phys. Rev. B **93**, 155128 (2016).
[31]   S. T. Weir, A. C. Mitchell, and W. J. Nellis, Phys. Rev. Lett. **76**, 1860 (1996).
[32]   V. E. Fortov *et al.*, Phys. Rev. Lett. **99**, 185001 (2007).
[33]   G. W. Collins *et al.*, Science **281**, 1178 (1998).
[34]   P. M. Celliers, G. W. Collins, L. B. D. Silva, D. M. Gold, R. Cauble, R. J. Wallace, M. E. Foord, and B. A. Hammel, Phys. Rev. lett. **84**, 5564 (2000).
[35]   G. W. Collins, P. M. Celliers, L. B. D. Silva, R. Cauble, D. M. Gold, M. E. Foord, N. C. Holmes, B. A. Hammel, and R. J. Wallace, Phys. Rev. Lett. **87**, 165504 1 (2001).
[36]   P. Loubeyre, S. Brygoo, J. Eggert, P. M. Celliers, D. K. Spaulding, J. R. Rygg, T. R. Boehly, G. W. Collins, and R. Jeanloz, Phys. Rev. B **86**, 144115 (9) (2012).
[37]   G. V. Boriskov, A. I. Bykov, R. I. Il'kaev, V. D. Selemir, G. V. Simakov, R. F. Trunin, V. D. Urlin, A. N. Shuikin, and W. J. Nellis, Phys. Rev. B **71**, 092104 (2005).
[38]   M. D. Knudson, M. P. Desjarlais, A. Becker, R. W. Lemke, K. R. Cochrane, M. E. Savage, D. E. Bliss, T. R. Mattsson, and R. Redmer, Science **348**, 1455 (2015).
[39]   I. F. Silvera and R. J. Wijngaarden, Rev. Sci. Instrum. **56**, 121 (1985).
[40]   A. Chijioke, W. J. Nellis, A. Soldatov, and I. F. Silvera, J. Appl. Phys. **98**, 114905 (2005).
[41]   Y. Akahama and H. Kawamura, Journal of Physics: Conference Series **215**, 012195 (2010).
[42]   R. C. Clay, J. Mcminis, J. M. McMahon, C. Pierleoni, D. M. Ceperley, and M. A. Morales, Phys Rev B **89**, 184106 (2014).
[43]   C. Pierleoni, M. A. Morales, G. Rillo, M. A. Strzhemechny, M. Holzmann, and D. M. Ceperley, PNAS **113**, 4953 (2016).
[44]   N. D. Drummond, B. Monserrat, J. H. Lloyd-Williams, P. L. Rios, C. J. Pickard, and R. J. Needs, Nature Comm. **6**, 7794 (2015).
[45]   S. Azadi, W. M. C. Foulkes, and T. D. Kühne, New Journal of Physics **15** (2013).
[46]   C. J. Pickard, M. Martinez-Canales, and R. J. Needs, Phys. Rev. B **85**, 214114 (2012).
[47]   I. B. Magdau and G. J. Ackland, Phys Rev B **87**, 174110 (2013).
[48]   H. Liu and Y. Ma, Phys. Rev. Lett. **110**, 025903 (2013).
[49]   R. T. Howie, I. B. Magdau, A. F. Goncharov, G. J. Ackland, and E. Gregoryanz, Phys. Rev. Lett. **113**, 175501 (2014).
[50]   C.S. Zha, Z. Liu, M. Ahart, R. Boehler, and R. J. Hemley, Phys. Rev. Lett. **110**, 217402 (2013).
[51]   C. S. Zha, R. E. Cohen, H. K. Mao, and R. J. Hemley, PNAS **111**, 4792 (2014).

[52] P. Loubeyre, F. Occelli, and P. Dumas, Phys Rev B **87**, 134101 (2013).
[53] M. I. Eremets, I. A. Troyan, P. Lerch, and A. Drozdov, High Pressure Research **33**, 377 (2013).
[54] R. Dias, O. Noked, and I. F. Silvera, PRL. **116**, 145501 (2016).
[55] Y. Akahama and H. Kawamura, J. Appl. Phys **96**, 3748 (2004).
[56] R. Dias, O. Noked, and I. F. Silvera, condensed matter arXiv:1605.05703 (2016).
[57] Y. Akahama, H. Kawamura, N. Hirao, Y. Ohishi, and K. Takemura, in *AIRAPT* (Journal of Physics: Conference Series, on-line http://iopscience.iop.org/1742-6596/215/1/012194, Tokyo, Japan, Journal of Physics: Conference Series, on-line http://iopscience.iop.org/1742-6596/215/1/012194, 2010), p. 012056.
[58] R. Howie, E. Gregoryanz, and A. F. Goncharov, J. Appl. Phys. **114**, 073505 (2013).